\documentclass[useAMS,usenatbib,usegraphicx]{mn2e}

\newcommand{\etal}{et al.}

\newcommand{\h}{1H~0707--495}

\title[Multiple X-ray reflection]
  {Multiple X-ray reflection from ionized slabs} 
\author[R.R. Ross, A.C. Fabian \& D.R. Ballantyne]
  {R.~R.~Ross$^1$\thanks{rross@holycross.edu}, A.~C.~Fabian$^2$
and D.~R.~Ballantyne$^2$\\
$^1$Physics Department, College of the Holy Cross, Worcester, MA 01610, USA \\
   $^2$Institute of Astronomy, Madingley Road, Cambridge CB3 0HA}

\pagerange{\pageref{firstpage}--\pageref{lastpage}}
\pubyear{2001}

\usepackage{times}

\begin{document}

\label{firstpage}

\maketitle

\begin{abstract}
Multiple reflection of X-rays may be important when an accretion disc
and its hot corona have a complicated geometry, or if returning
radiation due to gravitational light bending is important, or in
emission from a funnel such as proposed in some gamma-ray burst
models. We simulate the effects of multiple reflection by modifying
the boundary condition for an X-ray illuminated slab.  Multiple
reflection makes the soft X-ray spectrum steeper (softer) and
strengthens broad emission and absorption features, especially the
K-shell features of iron.  This may be important in explaining the
spectra of sources such as the Narrow-Line Seyfert 1 galaxy \h.
\end{abstract}


\section{Introduction}
The hard X-ray emission from luminous Seyfert 1 galaxies is often
modelled as due to an optically-thin corona above a dense accretion
disc. This model successfully accounts for the reflection component
often found in the spectra of these objects (Pounds \etal\ 1990). The dense
optically-thick accretion disc is often assumed to be relatively cold,
with iron only weakly ionized (e.g. George \& Fabian 1991; Matt, Perola 
\& Piro 1991; Magdziarz \& Zdziarski 1995). Sometimes however it may be
highly ionized by the irradiating X-rays, resulting in a more complex
spectrum (Ross \& Fabian 1993; \.{Z}ycki \etal\ 1994; Ross, Fabian \&
Young 1999; Nayakshin, Kazanas \& Kallman 2000; Ballantyne, Ross \&
Fabian 2001; R\'{o}\.{z}a\'{n}ska \etal\ 2002). Generally the disc is 
assumed to be flat.

Recent numerical simulations of luminous accretion discs (Turner,
Stone \& Sano 2002), particularly when radiation pressure is
important, suggest that they are clumpy, irregular and possibly
corrugated. In this case, if the corona lies close to the disc, the
solid angle subtended by the disc at the corona may be much higher
than $2\pi$, the value for a flat disc, and the reflection spectrum
itself may be reflected. This will certainly happen if the disc is
corrugated and the corona lies close to the bottom of corrugations.
The observer may then not see any direct emission from the corona but
only multiply-scattered flux.  (Rapid variability of an AGN requires
that any such corrugations be on a relatively small spatial scale.)

We attempt here to simulate multiply-scattered reflection spectra for
ionized discs in a simple generic manner using the code described by
Ross \& Fabian (1993). Multiply-scattered reflection spectra have been
invoked as one explanation (Fabian \etal\ 2002) for the remarkable
spectrum of the Narrow Line Seyfert~1 galaxy \h\ observed by
XMM-Newton (Boller \etal\ 2002). They might also be relevant if the
accretion disc extends well within 6 Schwarzschild radii so that
returning radiation due to light bending by the strong gravity of the
black hole is important (Cunningham 1975; Martocchia, Matt \& Karas
2002). Multiple reflection is expected naturally if the accretion flow
consists of clouds rather than a disc (e.g. Celotti, Fabian \& Rees
1992; Collin-Souffrin \etal\ 1996; Malzac 2001). Multiple reflection
could also be important in hypernova models for Gamma-ray bursts
(M\'{e}sz\'{a}ros 2001), in which the emission is surrounded by
ejecta.

We expect that multiple reflection will enhance both broad absorption
and emission features in the spectrum, relative to single reflection
spectra. The net effect is to produce a considerably steeper (softer)
spectrum with pronounced broad spectral features around the K-shell
emission of iron and other elements.

\section{The model and method}

As an idealized model, let us consider that the hard continuum
originates from close to the bottom of a well or hole. If the opening
at the top subtends an effective solid angle $\Omega$ as seen from the
bottom, then a fraction $\varepsilon=\Omega/2\pi$ of the emitted
radiation escapes directly; the rest is scattered, absorbed and
reflected by the material in the sides and bottom of the well. To a
crude approximation, this reflected flux has a similar probability of
escaping, with the rest restriking the surfaces within the well. We model 
the multiple reflection process by simply introducing a partially
reflecting mirror above a flat slab. A detailed model would involve
knowledge of the precise geometry envisaged, but we expect that our
simple model will give us a good qualitative guide to what the
observer would expect to detect. If $\varepsilon$ is small, an
observer may see only the multiply-reflected flux, and it is this
reprocessed radiation only that we discuss.

In detail, we treat a surface layer of gas with hydrogen number density
$n_{\rm H}=10^{15}\ {\rm cm}^{-3}$ and Thomson depth $\tau_{\rm
T}=10$. The outer surface is illuminated by ``direct'' radiation in
the form of a power law with photon index $\Gamma=2$ and with total
flux $F_0$ (from 1 eV to 100 keV) corresponding to a specified
ionization parameter
\begin{equation}
\xi = \frac{4\pi F_0}{n_{\rm H}}.
\end{equation}
The penetration of the illuminating radiation and the Compton-thick
transfer of the diffuse radiation produced via scattering or emission
within the gas are treated as described by Ross \& Fabian (1993).
At each step in the radiative
transfer calculation, the diffuse radiation emerging from the outer
surface is calculated as described by Foster, Fabian \& Ross (1986).
Only a fraction $\varepsilon$ of this emergent flux, however, is
allowed to escape to distant regions.   The remaining fraction
$(1-\varepsilon)$ of the emergent flux is assumed to be reflected back
onto the surface, so it is added to the ``direct'' radiation that
illuminates the surface.  The transfer calculation proceeds until the
radiation field relaxes to a steady state, at which point the portion
of the total emergent flux that actually escapes equals the total 
``direct'' illumination.

\begin{figure}
\includegraphics[width=0.34\textwidth,angle=270]{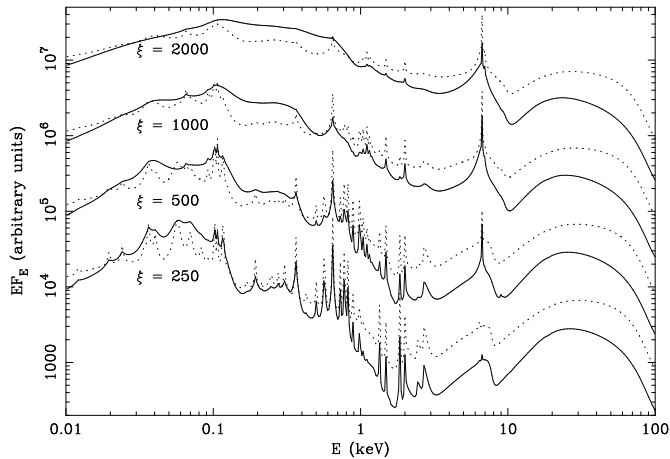}
\caption{Observed spectra resulting from multiple reflection with
escape fraction $\varepsilon=0.25$ (solid lines) and from a single
reflection (dotted lines) for four different values of the 
ionization parameter $\xi$.  Spectra for different values of $\xi$
have been offset vertically for clarity.}
\end{figure}
\begin{figure}
\includegraphics[width=0.34\textwidth,angle=270]{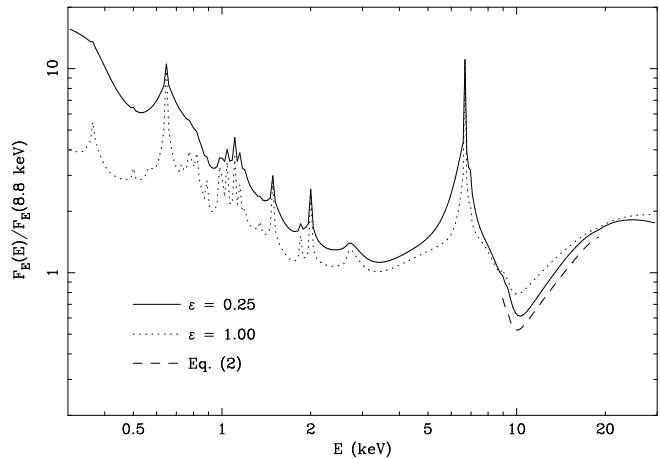}
\caption{Spectra for $\xi=1000$ renormalized by dividing the escaping
flux $F_E(E)$ by the corresponding value at $E$~= 8.8~keV.  The solid
line is the result for multiple reflection with $\varepsilon=0.25$, 
while the dotted line is the result for a single reflection  
($\varepsilon=1$).  The dashed line shows the approximate shape of the 
Fe K-absorption feature for $\varepsilon=0.25$ predicted using Eq.~(2).}
 
\end{figure}
\begin{figure}
\includegraphics[width=0.42\textwidth,angle=270]{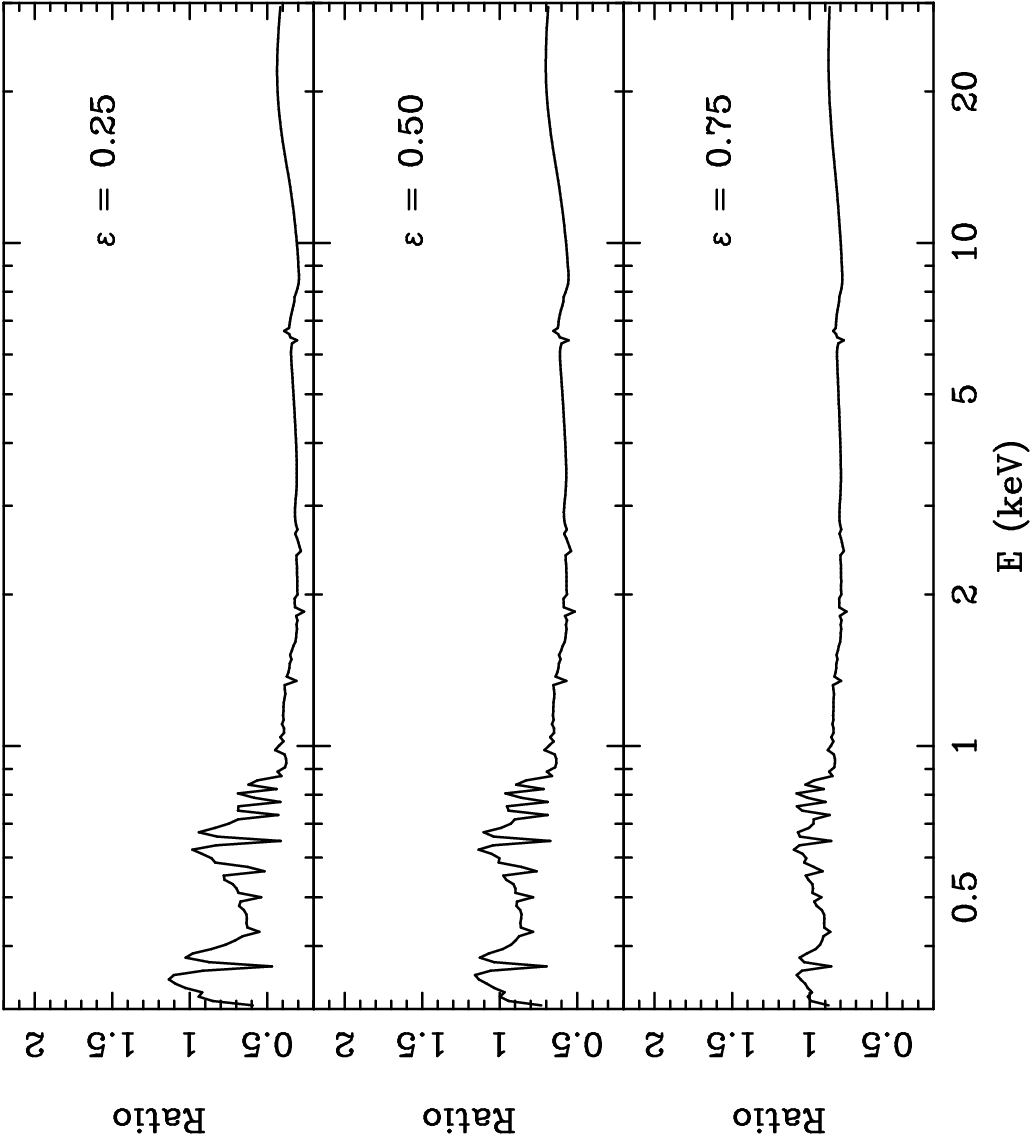}
\vskip0.25cm
\includegraphics[width=0.42\textwidth,angle=270]{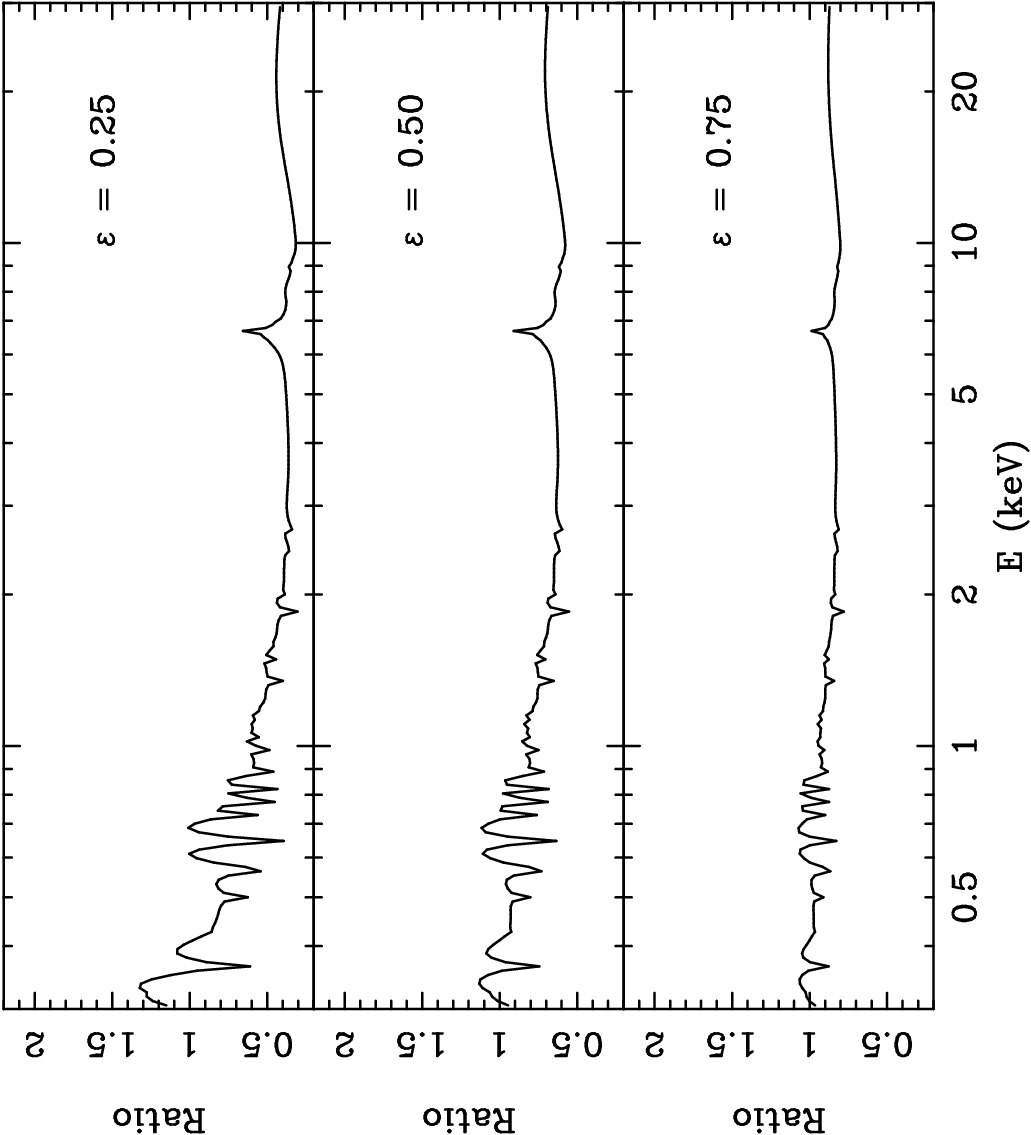}
\caption{Ratio of the observed flux after multiple reflection with
escape fraction $\varepsilon$ to the observed flux after a single
reflection.  The upper panel shows the results for $\xi=250$;  
the lower panel shows the results for $\xi=500$.}
\end{figure}
\begin{figure}
\includegraphics[width=0.42\textwidth,angle=270]{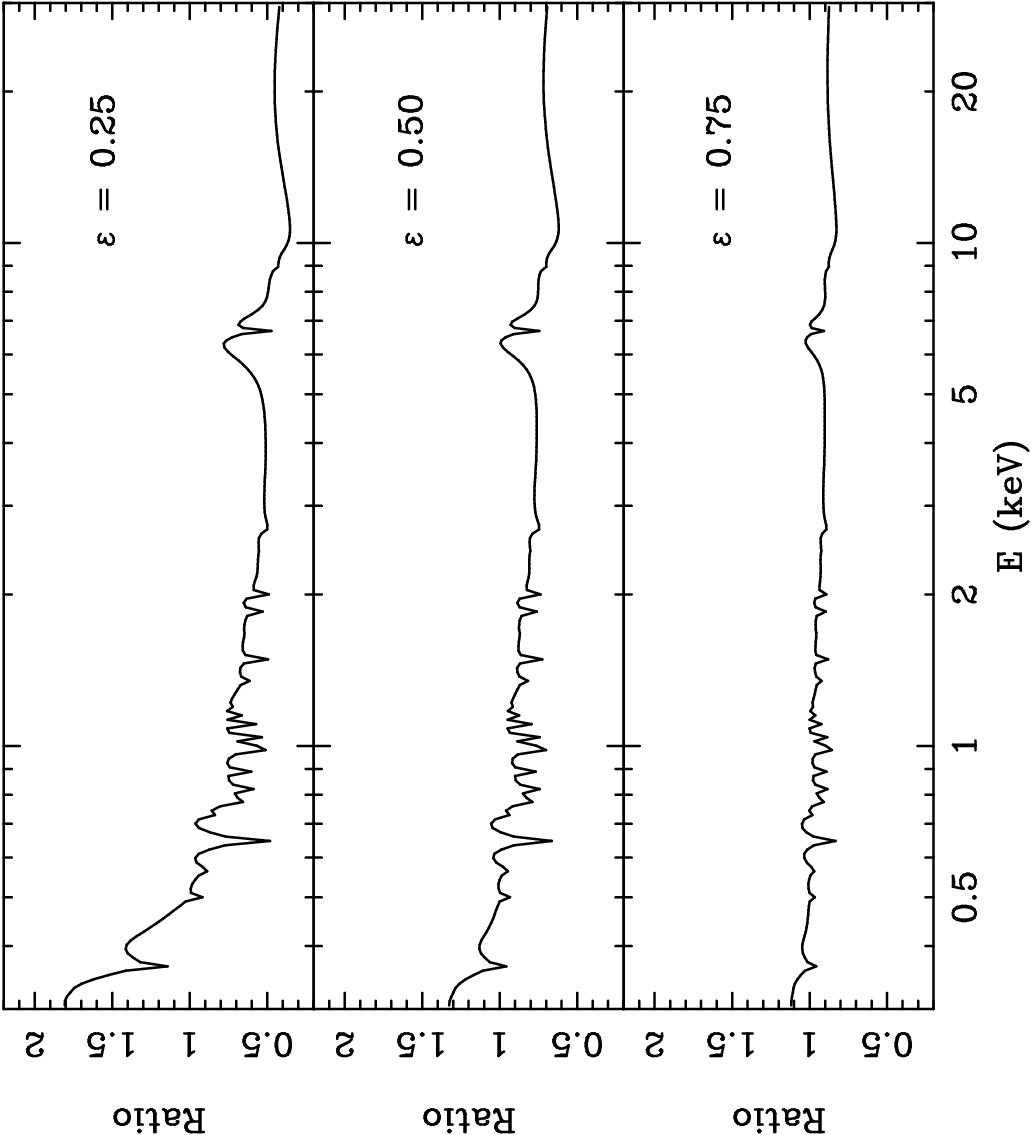}
\vskip0.25cm
\includegraphics[width=0.42\textwidth,angle=270]{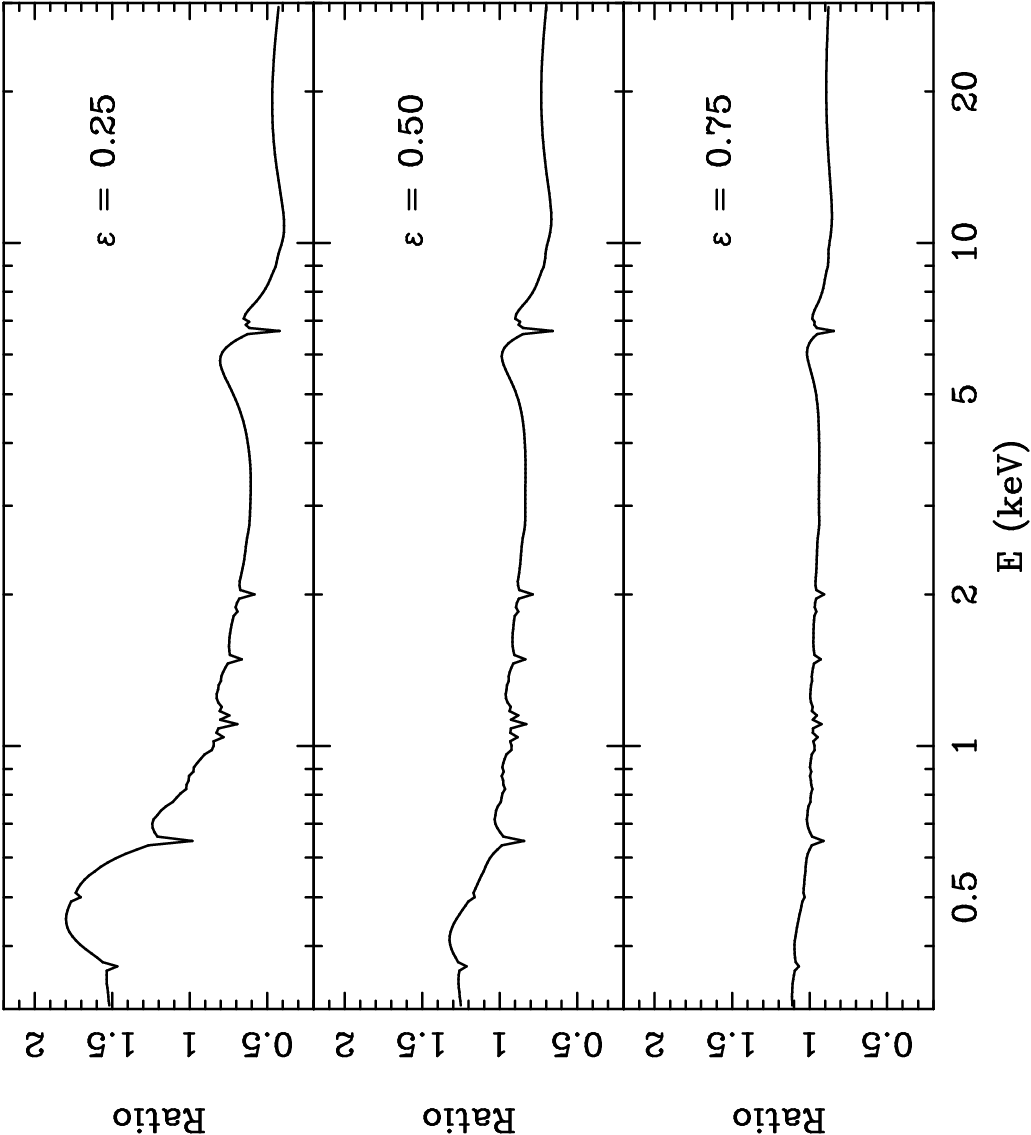}
\caption{Flux ratios as in Fig.~3, except for $\xi=1000$ (upper
panel) and for $\xi=2000$ (lower panel).}
\end{figure}
\begin{figure}
\includegraphics[width=0.34\textwidth,angle=270]{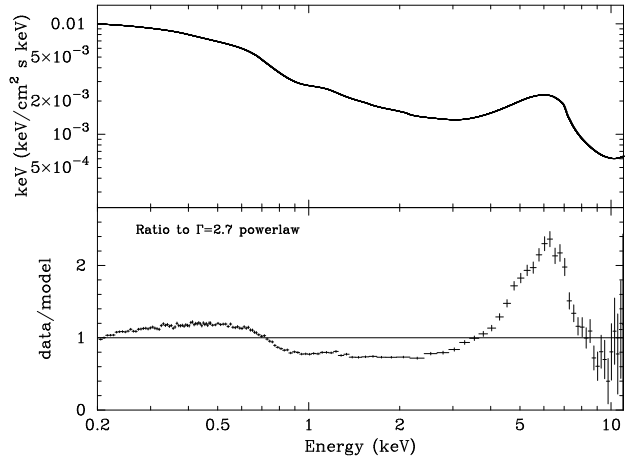}
\caption{Top: Model spectrum for $\varepsilon=0.25$ \& $\xi=2000$
relativistically blurred by a kernel appropriate (Laor 1991) for a
disk inclined at 30 deg to the line of sight and  extending from 3 to
100 gravitational radii (3--100$\times GM/c^2$). Bottom: Ratio of 
fake data obtained by simulating the above model for the EPIC pn CCD for
40~ks (and intensity appropriate for \h) to a a power-law model of
photon index 2.7.}
\end{figure}

\section{Results}

We have calculated multiple-reflection models with direct illumination
corresponding to $\xi=$ 250, 500, 1000 and 2000.  For each value of
the ionization parameter, models were produced with the emergent
radiation having escape fractions of $\varepsilon=$ 0.25, 0.50, 0.75
and 1.00.  Here $\varepsilon = 1$ corresponds to escape of all
emerging radiation with no reflection back onto the surface, and 
these results are similar to those of Ross, Fabian \& Young (1999).

Figure~1 shows the observed spectrum when only 25\% of the emergent
radiation actually escapes (and 75\% is reflected back) in comparison
with the spectrum without multiple reflection.  Multiple reflection
reduces the escaping flux for $E>1\ {\rm keV}$ by increasing the
probability that such photons will be absorbed before ultimately
escaping.  In addition, it changes the {\em shape} of the observed 
spectrum.  To make this clearer, Figure~2 shows the same two spectra 
for $\xi=1000$, but with each spectrum renormalized relative to the 
corresponding escaping flux at 8.8 keV (just below the K-edge of 
Fe~{\sc xxv}).  The multiple-reflection spectrum exhibits a relatively 
deeper Fe K-edge, a stronger and broader Comptonized Fe K$\alpha$ line, 
and a steeper spectrum in soft X-rays ($\sim {\rm 1\ keV}$).

An approximate test can be made of the accuracy of our method.  
Consider a photon energy for which the illuminated material is 
essentially a pure absorber and coherent scatterer---that is, there is 
no emission due to absorbed or scattered radiation at other photon 
energies.  Let $A$ be the albedo for a single reflection by the material.  
Then, if the probability of escape after each reflection is $\varepsilon$,
the net albedo produced by multiple reflection is
\begin{eqnarray}
A_{\rm net}&=&\varepsilon A + \varepsilon(1-\varepsilon)A^2 +
\varepsilon(1-\varepsilon)^2A^3 + \cdots \nonumber \\ 
&=&\frac{\varepsilon A}{1-(1-\varepsilon)A}
\end{eqnarray}
Fig.~2 shows the results of applying this formula to energies
around the K-edge of iron, where it is approximately valid, for 
$\varepsilon=0.25$.  Eq.~(2) predicts a slightly deeper
absorption feature than our multiple-reflection calculation
because it does not take into account incoherent Compton scattering
of higher-energy photons, which partially fills in the absorption
feature (see Ross, Weaver \& McCray 1978).

Figures~3 and~4 provide another way of visualizing the changes that
multiple reflection produce in the observed spectrum.  In each case,
the ratio of the flux escaping after multiple reflections to the flux
escaping after a single reflection is shown as a function of photon
energy.  On average, decreasing the escape fraction $\varepsilon$
decreases the entire spectrum above $\sim 1\ {\rm keV}$.  As
$\varepsilon$ decreases, however, the broad Fe K-absorption feature
decreases by a larger amount than the X-ray spectrum as a whole, so
that the {\em relative} strength of this feature is enhanced. On the
other hand, as $\varepsilon$ decreases, the broad, Comptonized Fe
K$\alpha$ emission feature is suppressed less than the X-ray spectrum
as a whole, so that this emission feature is also enhanced. At high
ionization parameters ($\xi=2000$ and 1000; Fig.~4), multiple
reflection suppresses the narrow central core of the Fe K$\alpha$
line, since there is a deep, highly-ionized surface layer that Compton
scatters line photons that reenter it.  At the lowest ionization
parameter ($\xi = 250$; Fig.~3), a decrease in $\varepsilon$ enhances
the Fe~{\sc xxv} line core at 6.7 keV and suppresses the ``neutral''
iron line at 6.4 keV, since multiple reflection increases the
effective ionization potential by increasing the total flux incident
on the surface.  In the soft X-ray region, the overall slope of the
ratio presented in Fig.~3 or~4 steepens as $\varepsilon$ decreases,
indicating that multiple reflection steepens the observed spectrum
compared to that resulting from a single reflection.

\section{Discussion}

Multiple X-ray reflection creates a steep soft X-ray spectrum and a
strong, broad, iron absorption/emission feature. The power lost from
the harder energies above $\sim 1\ {\rm keV}$ emerges at lower energies,
principally in the 0.05--1~keV band. Overall the spectra resemble
those from slightly higher ionization parameters and much higher
abundances (Ballantyne, Fabian \& Ross 2002). 

A relativistically blurred version of the spectrum for $\varepsilon=0.25$
and $\xi=2000$ is shown in Figure~5.  That has been used to simulate what 
would be observed with the XMM-Newton EPIC pn CCD, and this is also
displayed in Fig.~5 in the form of a ratio to an incident
power-law of photon index 2.7. The result compares well to the
spectrum of \h\ found by Boller \etal\ (2002). If we fit the simulated
spectrum from 2--10~keV with a power-law plus a broad gaussian line,
then we find that the best-fitting line has a central energy of
4.8~keV, a width of 1.5~keV and an equivalent width of about 6~keV.
Positive residuals remain over the 6--7 keV band. It is not our
intention to carry out a detailed fit to the data of \h, which would
require a comprehensive grid of models, but merely to confirm that
multiple reflection may be relevant there, as suggested by Fabian et
al. (2002). Since the spectrum of \h\ is unusual, multiple reflection 
may occur only occasionally among Narrow Line Seyfert 1 galaxies, and/or 
the primary continuum may be seen in addition to the reprocessed 
component most of the time.

Multiple reflection can be important when light bending is strong
enough to cause part of the reflected spectrum to return to the disk.
This occurs close to the centre of disks around maximally spinning
Kerr black holes (Martocchia \& Matt 1996; Dabrowski et al 1997;
Dabrowski \& Lasenby 2001; Martocchia, Matt \& Karas 2002). Strong
X-ray reflection has also been invoked (Vietri et al 2001; Ballantyne
\& Ramirez-Ruiz 2001) for the production of X-ray
lines seen in some gamma-ray bursts (e.g. Piro et al 2000). If the
emission is from a funnel geometry (Rees \& Meszaros 2000; McLaughlin
et al 2002) then the calculations presented here could be relevant.
Lower abundances are implied than obtained from single reflection
calculations.
 
\section{Acknowledgements}
The authors thank the anonymous referee for helpful comments.
ACF thanks the Royal Society for support, RRR acknowledges support
from the College of the Holy Cross, and DRB acknowledges financial
support from the Commonwealth Scholarship and Fellowship Plan and the
Natural Sciences and Engineering Council of Canada


\bsp 

\label{lastpage}

\end{document}